\documentclass[fleqn]{aa}
\usepackage[psamsfonts]{euscript}
\usepackage[psamsfonts]{amssymb}
\usepackage{amsmath}
\usepackage{graphicx}
\usepackage{natbib}

\newcommand{\nc}   {\newcommand}
\nc{\dsfrac}[2]{{\displaystyle\frac{#1}{#2}}}

\nc{\cm}    {\mathrm{cm}}
\nc{\ergs}  {\mathrm{erg}/\mathrm{gm\;s}}
\nc{\gcc}   {{\rm gm}~{\rm cm}^{-3}}
\nc{\K}     {\:\mathrm{K}}
\nc{\kms}   {{\rm km}~{\rm s}^{-1}}

\nc{\aT}    {a _\mathrm{T}}
\nc{\CCa}  {\EuScript{C}_0}
\nc{\CCb}  {\EuScript{C}_1}
\nc{\CCc}  {\EuScript{C}_2}
\nc{\Ee}    {E_\mathrm{e}}
\nc{\Eem}   {\widetilde{E}_\mathrm{e}}
\nc{\Eexm}  {\widetilde{E}_\mathrm{ex}}
\nc{\Eim}   {\widetilde{E}_\mathrm{I}}
\nc{\ER}    {E_\mathrm{R}}
\nc{\FL}    {F_\mathrm{Ly}}
\nc{\FR}    {F_\mathrm{R}}
\nc{\FRa}   {F_\mathrm{R,1}}
\nc{\FRn}   {F_\mathrm{R,N}}
\nc{\FFR}   {\mathbf{F}_\mathrm{R}}
\nc{\Fe}    {F_\mathrm{e}}
\nc{\FFe}   {\mathbf{F}_\mathrm{e}}
\nc{\Fsat}  {q_\mathrm{sat}}
\nc{\Jcal}  {\EuScript{J}}
\nc{\Ke}    {\EuScript{K}_\mathrm{e}}
\nc{\me}    {m_\mathrm{e}}
\nc{\mfp}   {l_\mathrm{e}}
\nc{\mrc}   {\mathrm{c}}
\nc{\mrr}   {\mathrm{r}}
\nc{\nH}    {n_\mathrm{H}}
\nc{\nel}   {n_\mathrm{e}}
\nc{\Pe}    {P_\mathrm{e}}
\nc{\Pg}    {P_\mathrm{g}}
\nc{\PR}    {P_\mathrm{R}}
\nc{\Qelc}  {Q_\mathrm{elc}}
\nc{\Qinc}  {Q_\mathrm{inc}}
\nc{\gyr}   {r_\mathrm{B}}
\nc{\Ta}    {T_a}
\nc{\Te}    {T_\mathrm{e}}
\nc{\Xca}   {X_{\mathrm{c}a}}
\nc{\Xcb}   {X_{\mathrm{c}b}}
\nc{\xH}    {x_\mathrm{H}}

\nc{\divFe} {\nabla\cdot\FFe}
\nc{\divFR} {\nabla\cdot\FFR}
\nc{\FLj}   {F_\mathrm{Ly,\Jcal}}
\nc{\FRj}   {F_\mathrm{R,\Jcal}}

\nc{\aap}   {A\&A}
\nc{\apj}   {ApJ}

\begin{document}

\title{The structure of radiative shock waves}
\subtitle{IV. Effects of electron thermal conduction}

\author{Yu.A. Fadeyev \inst{1} \and
        H. Le Coroller\inst{2} \and
        D. Gillet\inst{2}}
\offprints{D. Gillet, \email{gillet@obs-hp.fr}}
\institute{Institute for Astronomy of the Russian Academy of Sciences,
 Pyatnitskaya 48, 109017 Moscow, Russia \and
 Observatoire de Haute-Provence - CNRS, F-04870
 Saint-Michel l'Observatoire, France}
\date{Received \hbox to 24pt {} September 2001 / Accepted }

\abstract{
We consider the structure of steady--state radiative shock waves propagating
in the partially ionized hydrogen gas with density $\rho_1 = 10^{-10}~\gcc$
and temperature $3000\K\le T_1\le 8000\K$.
The radiative shock wave models with electron thermal conduction in the
vicinity of the viscous jump are compared with pure radiative models.
The threshold shock wave velocity above of which effects of electron thermal
conduction become perceptible is found to be of $U_1^*\approx 70~\kms$ and
corresponds to the upstream Mach numbers from $M_1\approx 6$ at $T_1=8000\K$
to $M_1\approx 11$ at $T_1=3000\K$.
In shocks with efficient electron heat conduction more than a half of hydrogen
atoms is ionized in the radiative precursor, whereas behind the
viscous jump the hydrogen gas undergoes the full ionization.
The existence of the electron heat conduction precursor leads to the
enhancement of the Lyman continuum flux trapped in the surroundings of the
discontinuous jump.
As a result, the partially ionized hydrogen gas of the radiative
precursor undergoes an additional ionization ($\delta\xH\lesssim 5\%$),
whereas the total radiave flux emerging from the shock wave
increases by $10\%\le\delta(\FR)\le 25\%$
for $70~\kms\le U_1 \le 85~\kms$.
\keywords{Shock waves -- Hydrodynamics -- Radiative transfer -- Stellar atmospheres}}

\titlerunning{The structure of radiative shock waves. IV}
\authorrunning{Yu.A. Fadeyev et al.}
\maketitle

\section{Introduction}

In our previous papers
\citep[][]{1998A&A...333..687F,2000A&A...354..349F,2001A&A...368..901F},
hereinafter referred to as Papers~I -- III,
we presented the studies of steady--state radiative shock waves propagating
in the partially ionized hydrogen gas with properties that are
typical for atmospheres of pulsating late--type stars.
The models were considered in terms of the self--consistent solution of the
equations of fluid dynamics, radiation transfer and rate equations for the
hydrogen atom and provide the reliable estimates of radiative energy losses
of the shock wave.
In these studies we adopted that at the viscous jump which is treated
as an infinitesimally thin discontinuity the electron gas compresses
adiabatically.
Such an assumption is justified by the very low rate of energy exchange
between protons and electrons within the viscous jump and thereby
allows us to determine the postshock electron temperature from the
simple adiabatic relation.
However, it is known that the characteristic length scale of electron thermal
conduction might be comparable with the thickness of the postshock relaxation
zone (the length of the temperature equilibration zone) and,
therefore, might affect perceptibly the spatial distribution of hydrodynamic
variables at least in the vicinity of the viscous jump
\citep[see, for discussion,][]{1967pswh.book.....Z,1984frh..book.....M}.

Effects of electron thermal conduction in two--temperature shock
waves propagating in partially ionized helium and argon with
temperatures and densities close to those of stellar atmospheres
were investigated using both hydrodynamic and kinetic approaches
by \cite{Grewal:Talbot:1963,Jaffrin:1965,Lu:Huang:1974}.
Solution of the more general problem involving the radiation transfer
was considered by \cite{Vinolo:Clarke:1973}.
The conspicuous result of these calculations is that the authors
demonstrated the existence of the
zone of the elevated electron temperature ahead the viscous jump
appearing due to the high thermal conductivity of the electron gas.
For the hydrogen gas effects of electron thermal conduction
were considered only in the limit of full ionization at Mach numbers
as high as $M_1\approx 8$ \citep{1993PhFlB...5.3182V,1995PhPl....2.1412V}.
This study, unfortunately, was confined to the problem of inertial
confinement fusion for the extremely high temperature and number density of
electrons ($T = 10^8\K$, $\nel = 10^{22}~\cm^{-3}$).

By now, the scarce studies of astrophysical radiative shock waves with 
electron thermal conduction in two--temperature gases were confined to the
interstellar medium \citep{1989ApJ...336..979B,1990ApJ...348..169B}
and accreting white dwarfs \citep{1987ApJ...313..298I}.
In these works the heat conduction was found to substantially
affect the structure of the postshock relaxation zone for
shocks with velocities ranged from $70~\kms$ to $170~\kms$.
In atmospheres of late--type pulsating stars the gas density
is many orders of magnitude higher in comparison with
that of the interstellar medium, whereas the velocities of
shocks does not exceed $100~\kms$, so that
the role of electron heat conduction in radiative losses
remains highly uncertain.
In this paper we compute the models of steady--state radiative shock waves
propagating through the partially ionized hydrogen gas with properties
typical for atmospheres of pulsating late--type stars
and compare the pure radiative shock wave models with those
in which electron the thermal conduction is taken into account.

As in our previous papers we assume that the ambient unperturbed medium
is homogeneous and effects of magnetic fields are negligible.
Indeed, the magnetic field becomes important when the electron
mean free path $\mfp$ is larger than the gyromagnetic radius $\gyr$.
For the gas density considered in our study ($\rho = 10^{-10}~\gcc$)
the condition $\mfp\gg\gyr$ fulfills for $B\gg 3$~Gauss.
Furthermore, the magnetic pressure $B^2/8\pi$ becomes comparable
with the gas pressure for $B > 25$~Gauss.
On the other hand, the strength of the magnetic field in late--type giants
is $B\ll 1$~Gauss \citep{1982A&A...105..133B}.
Thus, effects of magnetic fields can be ignored without any loss
of accuracy.
We expect that results of our calculations can be applied to shock phenomena
observed in atmospheres of radially pulsating giants and supergiants
such as W~Vir, RV~Tau and Mira type variables.

In order to specify the model we use three parameters
determining the structure of the steady--state radiative shock wave.
These are the density $\rho_1$ and the temperature $T_1$ of the
ambient hydrogen gas and the upstream gas flow
velocity $U_1$.

\section{Equations for thermal conduction}

Detailed discussion of the basic equations used for calculation of the
structure of radiative shock waves is given in our earlier papers
and in this section we only consider the equations involving the terms
responsible for electron thermal conduction.
Eq.~(20) of Paper~II describes
the change of the translational specific energy of the electron gas
\begin{equation}
\Eem =\frac{3}{2}\frac{\nel}{\rho} k\Te
\end{equation}
and is written now as
\begin{eqnarray}
\label{deedt}
\dsfrac{d\Eem}{dt} =
&-& \Pe\dsfrac{dV}{dt} + \Qelc - \dsfrac{d\Eim}{dt} - \dsfrac{d\Eexm}{dt}
- \dsfrac{1}{\rho}\dsfrac{d\ER}{dt} -
\nonumber\\
&-& \dsfrac{1}{\rho}\divFR - \dsfrac{1}{\rho}\divFe .
\end{eqnarray}
Here $\nel$ is the number density of free electrons,
$\Pe = \nel k\Te$ is the partial pressure of the electron gas,
$V=1/\rho$ is the specific volume,
$\Qelc$ is the rate of energy gain by electrons in elastic collisions
with neutral hydrogen atoms and hydrogen ions,
$\Eim$ and $\Eexm$ are the ionization energy and the excitation energy
per unit mass,
$\ER$ and $\FFR$ are the radiation energy density and the total radiation flux,
respectively,
\begin{equation}
\label{fe}
\FFe = - \Ke\nabla\Te
\end{equation}
is the electron heat conductive flux.
The thermal conductivity $\Ke$ was calculated using the fitting formulae
\citep{1977A&A....60..413N} taking into account effects of partial
ionization and converging with those given by \cite{1962pfig.book.....S}
for the fully ionized hydrogen.

Equation (\ref{fe}) is valid until the mean free path of electrons is smaller
than the characteristic temperature scale and for the too steep temperature
gradient it becomes inapplicable due to effects of saturation of the heat flux.
According to \cite{1977ApJ...211..135C} the saturated heat flux $\Fsat$
is proportional to the product of the thermal energy of electrons
and their mean thermal velocity, that is,
\begin{equation}
\label{fsat}
\Fsat = f\left(\frac{2k\Te}{\pi\me}\right)^{1/2}\nel k\Te .
\end{equation}
Unfortunately, the value of the flux limit factor $f$
for the partially ionized hydrogen gas is uncertain.
\cite{1977ApJ...211..135C} adopted $f = 0.4$, whereas
\cite{1989ApJ...336..979B} presented arguments in favour of $f\sim 0.1$.
According to more recent kinetic calculations \citep{1995PhPl....2.1412V}
the flux limit factor is $f\approx 0.3$ and it is the value that we used
in our study.

The presence of the divergence of the conductive flux in Eq.~(\ref{deedt})
increases the order of this equation and leads to the much stronger
sensitivity of the resulting system of ordinary differential equations
(see Eqs.~(17) -- (21) of Paper~II) to the round--off errors.
To keep the first order of Eq.~(\ref{deedt}) we calculated $\divFe$
each iteration together with solution of the radiation transfer equation
for the given spatial distribution of hydrodynamic variables.
For the first iteration we replaced the spatial distribution of the
electron temperature $\Te$ obtained from the pure radiative model
and having the jump at the discontinuity by the arbitrary function
smoothing the run of $\Te$ within $\sim 100~\cm$
in both sides of the discontinuous jump.
After the series of trial calculations we found that the final solution
does not depend on
the initial distribution
of $\Te$ since the region of the effective energy transfer by
electron thermal conduction is by an order of magnitude narrower.
The shock wave models were represented by
$2000\le N\le 8000$ cells with
$\sim 10^3$ cells within the thermal conduction region.

The divergence of the conductive flux was computed using the second--order
finite--difference formula
\begin{equation}
\begin{array}{l}
\left(\divFe\right)_{j-1/2} =
-\dsfrac{{\Ke}_{j-1}}{\Delta X_{j-1}}\dsfrac{{\Te}_{j-3/2}}{\Delta X_{j-1/2}} +
\\[16pt]
+ \left(
\dsfrac{{\Ke}_{j-1}}{\Delta X_{j-1}} + \dsfrac{{\Ke}_j}{\Delta X_j}
\right)\dsfrac{{\Te}_{j-1/2}}{\Delta X_{j-1/2}} -
\dsfrac{{\Ke}_j}{\Delta X_j}\dsfrac{{\Te}_{j+1/2}}{\Delta X_{j-1/2}} ,
\end{array}
\end{equation}
where
$\Delta X_j = \frac{1}{2}\left(\Delta X_{j-1/2} + \Delta X_{j+1/2}\right)$
and $\Delta X_{j-1/2} = X_j - X_{j-1}$ are the space intervals.
All thermodynamic quantities are defined at cell centers having
half--integer subscripts and the conductivity coefficient at cell
interfaces is determined as the mass--weighted average:
\begin{equation}
{\Ke}_j =
\frac{{\Ke}_{j-1/2}\Delta m_{j-1/2} + {\Ke}_{j+1/2}\Delta m_{j+1/2}}
{\Delta m_{j-1/2} + \Delta m_{j+1/2}} ,
\end{equation}
where $\Delta m_{j-1/2} = \rho_{j-1/2}\Delta X_{j-1/2}$
is the mass contained in the cell $\left[X_{j-1}, X_j\right]$.
Thus, the divergence of the conductive flux in Eq.~(\ref{deedt})
was considered explicitly and during integration of the ordinary
differential equations was calculated using the nonlinear interpolation.

The discontinuity is assumed to locate at the $\Jcal$--th cell interface
with space coordinate $X_\Jcal = 0$, so that coordinates of cell centers just
ahead and just behind the discontinuity are
$X_{\Jcal - 1/2}$ and $X_{\Jcal + 1/2}$.
For the sake of convenience the quantities defined at these cell centers
are denoted by superscripts minus and plus, respectively.
In all models considered in our study the size of the
central cell was set equal to
$\Delta X_\Jcal = X_{\Jcal + 1/2} - X_{\Jcal - 1/2} = 0.1~\cm$.

Solution of the preshock initial--value problem is ended at the
$(\Jcal - 1/2)$--th cell center and in order to obtain the initial
conditions for the postshock integration we use the Rankine--Hugoniot
equations relating quantities at both sides of the discontinuity at
cell centers $X_{\Jcal - 1/2}$ and $X_{\Jcal + 1/2}$.
For the radiative shock wave with electron heat conduction the
Rankine--Hugoniot equations are written as
\begin{eqnarray}
\label{RHa}
&&\rho U = \CCa \equiv\dot m ,\\
\label{RHb}
&&\dot mU + \Pg + \PR = \CCb, \\
\label{RHc}
&&\frac{1}{2}\dot mU^2 + \dot mh + \Fe + \FR + U\left(\ER + \PR\right) = \CCc ,
\end{eqnarray}
where $h$ is the specific enthalpy (see Eq.~(7) of Paper~II),
$\Pg$ is the gas pressure, $\PR$ is the radiation pressure,
$\CCa$, $\CCb$ and $\CCc$ are the mass, momentum and energy fluxes.
Solution of the Rankine--Hugoniot relations gives 
the compression ratio across the discontinuous jump
$\eta = \rho^+/\rho^-$ as the root of the quadratic equation
\begin{equation}
A\eta^2 - B\eta + C = 0 ,
\end{equation}
where
\begin{eqnarray}
A &=& \left(\aT^-\right)^2 + \frac{1}{5}\left(U^-\right)^2 -
\frac{2}{5}\frac{\Fe^+ + \FR^+ - \Fe^- - \FR^-}{\dot m} +
\nonumber\\
&+&\frac{2}{5}\frac{\ER^- + \PR^-}{\rho^-} ,
\\
B &=& \left(\aT^-\right)^2 + \left(U^-\right)^2 +
\frac{2}{5}\frac{\ER^+ + \PR^+}{\rho^-} -
\frac{\PR^+ - \PR^-}{\rho^-} ,
\\
C &=& \frac{4}{5}\left(U^-\right)^2 ,
\end{eqnarray}
$\aT = \left(\Pg/\rho\right)^{1/2}$ is the isothermal sound speed.

In contrast to our earlier studies where we assumed that the electron
gas compresses at the viscous jump adiabatically
in the present paper
we employed the jump condition for the electron temperature written as
\begin{equation}
\Te^+ = \Te^- - \frac{{\Fe}_\Jcal}{{\Ke}_\Jcal}\Delta X_\Jcal  .
\end{equation}
The postshock temperature of heavy particles is obtained from solution
of Eqs.~(\ref{RHa} -- \ref{RHc}) and is given by
\begin{eqnarray}
\Ta^+ &=& \Ta^- - \frac{\nel^-}{\nH^-}\left(\Te^+ - \Te^-\right)
+ \frac{1}{5}\frac{\dot mU^-}{\nH^- k}\frac{\eta^2 - 1}{\eta^2} -
\nonumber\\
& - & \frac{2}{5}\frac{\Fe^+ + \FR^+ - \Fe^- - \FR^-}{\nH^- kU^-} -
\nonumber\\
& - & \frac{2}{5}
\frac{\left(\ER^+ + \PR^+\right)\eta^{-1} - \left(\ER^- + \PR^-\right)}
{\nH^-k} ,
\end{eqnarray}
where $\nH$ is the number density of hydrogen atoms.

\section{The structure of shocks with heat conduction}

In solution of the initial--value problem for the fluid dynamics
and rate equations we considered the system of ordinary
differential equations described in Paper~II.
The only exception is that within the interval
$\Xca=-100~\cm\le X\le\Xcb=100~\cm$ surrounding the discontinuous
jump the equation for $\Ee$ (Eq.~(20) of Paper~II) is replaced
by Eq.~(\ref{deedt}).
The large size of the interval with heat conduction taken into account
was used in order
to diminish the role of boundary conditions $\Fe(\Xca)=\Fe(\Xcb)=0$
in calculations of the divergence of conductive flux $\nabla\cdot\FFe$.

\begin{figure}
\resizebox{\hsize}{!}{\includegraphics{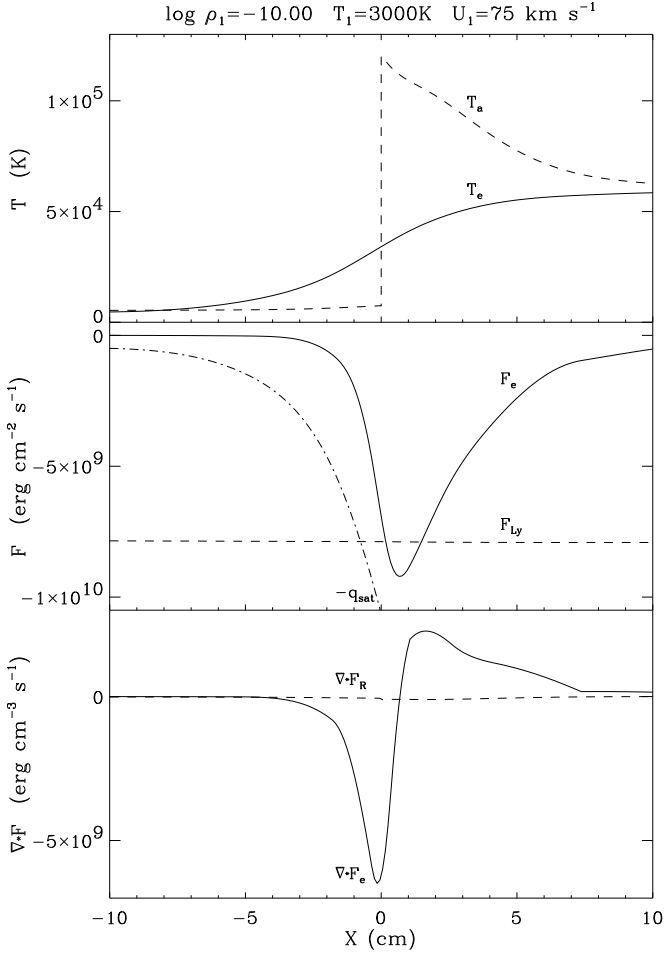}}
\caption{The vicinity of the discontinuous jump located at $X=0$
of the shock wave model
with $\rho_1=10^{-10}~\gcc$, $T_1=3000\K$ and $U_1=75~\kms$.
Upper panel: the electron temperature $\Te$ (solid line)
and the temperature of heavy particles $\Ta$ (dashed line).
Middle panel: the heat conductive flux $\Fe$ (solid line),
the Lyman continuum flux $\FL$ (dashed line )and
the saturated heat flux $\Fsat$ (dot--dashed line).
Lower panel: the divergence of the heat conductive flux $\nabla\cdot\FFe$
(solid line) and the divergence of the total radiative flux
$\nabla\cdot\FFR$ (dashed line).}
\label{fig1}
\end{figure}

The vicinity of the discontinuous jump of the shock wave model with
$\rho_1=10^{-10}~\gcc$, $T_1=3000\K$ and $U_1=75~\kms$ is shown
in Fig.~\ref{fig1} where as a function of $X$ are plotted
the electron temperature $\Te$ and the temperature of heavy particles $\Ta$
(upper panel),
the heat conductive flux $\Fe$, the Lyman continuum flux $\FL$ and
the saturated heat flux $\Fsat$ (middle panel),
the divergence of the heat conductive flux $\nabla\cdot\FFe$ and
the divergence of the total radiative flux $\nabla\cdot\FFR$
(lower panel).
It should be noted that
in the framework of our model the space coordinate $X$ increases while
the gas element passes through the shock wave and in such a notation
the upstream and downstream fluxes are negative and positive, respectively.

In the unperturbed hydrogen gas with density $\rho_1=10^{-10}~\gcc$ and
the temperature of $3000\K\le T_1\le 6000\K$
the length of the radiative precursor is $\sim 10^4~\cm$ and
at the upstream gas flow velocity $U_1 = 75~\kms$
the temperature growth due to absorption of the Lyman continuum
radiation emerging from the postshock region is
$(\Delta T)_\mrr\approx 2000\K$.
As is seen from Fig.~\ref{fig1} the length of the region of efficient
heat conduction surrounding the discontinuous jump is of $\sim 10$~cm.
Within such a narrow interval the divergence of the heat conductive flux
exceeds the divergence of radiative flux by almost two orders of
magnitude, so that the heating and cooling of gas are due to
the electron heat conduction and the role of radiation is negligible.

For the model represented in Fig.~\ref{fig1}
the electron temperature growth within the conductive precursor is
of $(\Delta\Te)_\mrc\approx 3\cdot 10^4\K$.
A small fraction of the translational energy of the electron gas
is lost in elastic collisions with hydrogen ions and the temperature of
heavy particles increases within the conductive precursor by
$(\Delta\Ta)_\mrc\approx 2000\K$.

The heat conductive flux $\Fe$ does not exceed the total radiative flux
and its maximum value is only comparable with the Lyman flux $\FL$
(see the middle panel of Fig.~\ref{fig1}) which is of $\approx 40$
percent of the total radiative flux $\FR$.
As is seen from Fig.~\ref{fig1} effects of saturation
can play perceptible role only in the conductive precursor just ahead
the discontinuous jump,
whereas behind the discontinuity $\Fsat\gg|\Fe|$ because of
higher density and higher temperature of the gas.

\begin{figure}
\resizebox{\hsize}{!}{\includegraphics{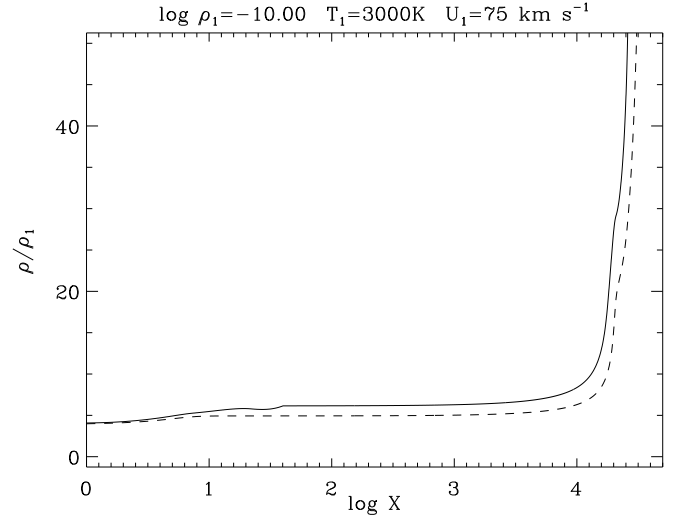}}
\caption{The postshock compression ratio $\rho/\rho_1$ in the shock wave
model with electron heat conduction (solid line) and in the pure radiative
model (dashed line).}
\label{fig2}
\end{figure}

The growth of the preshock electron temperature due to conductive heating
is accompanied by increase of the gas density.
For example, in the pure radiative model with
$\rho_1=10^{-10}~\gcc$, $T_1=3000\K$ and $U_1=75~\kms$
the preshock compression ratio is $(\rho^-/\rho_1)_\mrr=1.008$,
whereas in the model with heat conduction the preshock compression ratio
is $(\rho^-/\rho_1)_\mrc=1.04$.
The difference in compression ratios becomes more significant
in the postshock region and increases with increasing distance from
the discontinuous jump.
For example, at the boundary of the thermalization zone
($\log X\approx 1$)
the ratio of gas densities evaluated for the conductive and
pure radiative models is
$\rho_\mrc/\rho_\mrr\approx 1.1$,
whereas at the boundary of the recombination zone ($\log X\approx 4.5$)
this ratio is as high as $\rho_\mrc/\rho_\mrr\approx 1.6$.
Two plots of the postshock compression ratios $\rho/\rho_1$
for the conductive and radiative shock wave models are shown in Fig.~\ref{fig2}.

\begin{figure}
\resizebox{\hsize}{!}{\includegraphics{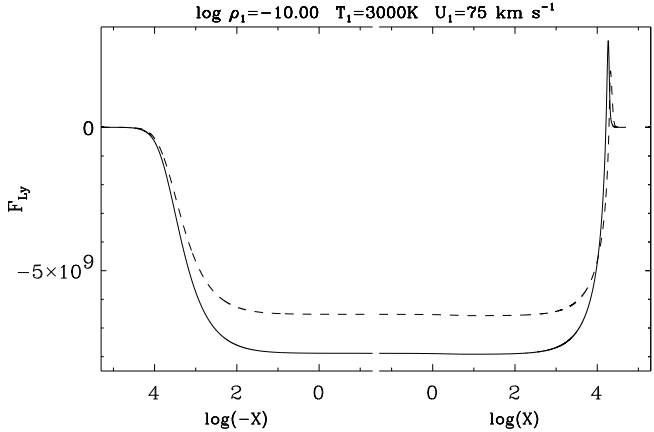}}
\caption{The Lyman continuum flux in the conductive (solid line)
and pure radiative (dashed line) shock waves.
The left and the right panels represent the preshock and
postshock regions, respectively.}
\label{fig3}
\end{figure}

The stronger compression in shock waves with heat conduction
implies that relaxation processes are somewhat faster in comparison with
those in pure radiative shocks.
Such a difference is due to the fact that existence of the
preshock conductive precursor leads to the elevation of the radiation
energy density within the whole area of the trapped Lyman continuum
radiation.
This area is confined by the layers of hydrogen ionization in the
preshock region (the radiative precursor) and by the layers
of hydrogen recombination behind the discontinuous jump.
The enhancement of the radiation energy density is illustrated in
Fig.~\ref{fig3} where we plot the Lyman continuum flux
as a fuction of $X$ for pure radiative and for the conductive model.
For the sake of convenience we use the logarithmic scales
for the both preshock and postshock regions that are represented by the left
and by the right panels, respectively.

\begin{figure}
\resizebox{\hsize}{!}{\includegraphics{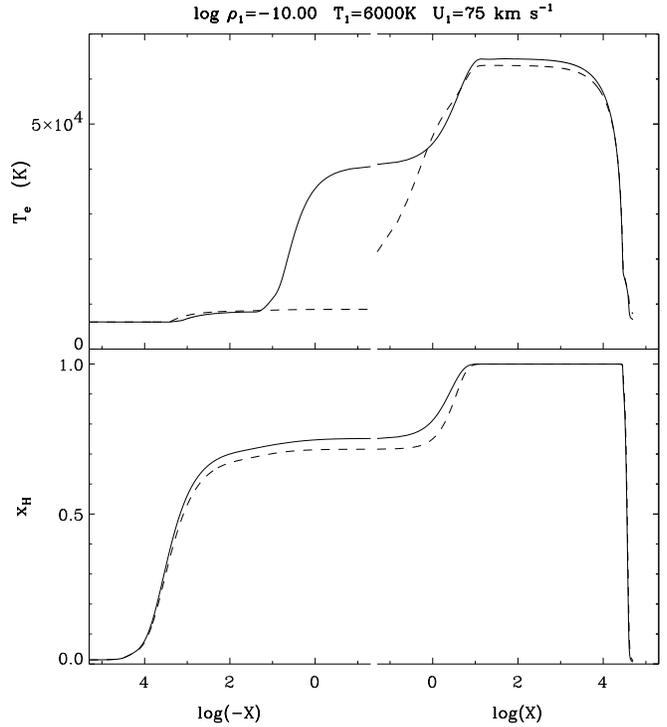}}
\caption{ The electron temperature $\Te$ (upper panel) and
the hydrogen ionization degree $\xH$ (lower panel).
In solid and dashed lines are represented the conductive and pure radiative
models.
The meaning of the left and the right panels is the same as in
Fig.~\ref{fig3}.}
\label{fig4}
\end{figure}

As was shown in our earlier papers (see, for example, Paper~I and Paper~II)
the populations of atomic levels are governed by radiative transitions
because the rates of corresponding collisional
transitions are several orders of magnitude smaller
throughout the whole shock wave.
Thus, elevation of the radiation energy density in the Lyman
continuum leads to the larger preshock hydrogen ionization degree
and at the same time to the faster ionization (if the hydrogen gas
is still partially ionized in the radiative precursor) in the postshock region.
This effect is illustrated in Fig.~\ref{fig4}
representing the spatial distributions of the electron temperature $\Te$
and the hydrogen ionization degree $\xH$ for the both conductive and
pure radiative shock wave models.

The efficiency of electron heat conduction in radiative shock waves depends on
the hydrogen ionization degree ahead the discontinuous jump as well as
on the electron temperature behind the discontinuous jump.
Obviously, both these quantities increase with increasing upstream gas flow
velocity $U_1$ though the preshock hydrogen ionization degree
depends also on the temperature $T_1$ of the unperturbed ambient gas.
In order to determine the threshold upstream velocity $U_1^*$ for electron
thermal conduction we computed a number of
models with different values of $T_1$ and $U_1$ but for the fixed value
of the unperturbed gas density $\rho_1 = 10^{-10}~\gcc$.
Results of these computations are summarized in Table~\ref{table}
where the columns headed by $U_1$, $T_1$ and $M_1$ give the upstream gas flow
velocity in $\kms$, the temperature of the unperturbed hydrogen gas and
the upstream Mach number.
The column headed by $\xH^-$ gives the hydrogen ionization
degree just ahead the discontinuous jump of the shock wave model with
electron heat conduction and the column headed by $\delta(\xH^-)$ gives
the raise of the preshock hydrogen ionization degree in comparison
with the pure radiative model.
We found that the preshock raise of the hydrogen ionization degree
due to effects of the electron heat conduction is
$\delta(\xH^-)\lesssim 0.05$.
Obviously, $\delta(\xH^-) = 0$ not only in the weak shocks with
negligible heat conduction but also
in strong shocks with full ionization of the preshock hydrogen gas.
In the latter case the excess of the radiation energy density in the Lyman
continuum leads to the stronger radiative heating of the
preshock gas.

In other to evaluate the role of electron thermal conduction in the structure
of radiative shock waves we compare the preshock gas density $\rho^-$,
the Lyman continuum flux at the discontinuous jump $\FLj$ and the total
radiative flux emerging from the both boundaries of the shock wave model
$\FR=\frac{1}{2}(\FRa+\FRn)$ with corresponding quantities evaluated
for the pure radiative models.
To this end in last three columns of Table~\ref{table} we list the ratios
$\varepsilon(\rho^-) = (\rho^-)_\mrc/(\rho^-)_\mrr$,
$\varepsilon(\FLj) = (\FLj)_\mrc/(\FLj)_\mrr$ and
$\varepsilon(\FR) = (\FR)_\mrc/(\FR)_\mrr$.

\begin{table}
\caption{Properties of radiative shock waves with electron heat conduction
at $\rho_1 = 10^{-10}~\gcc$.}
\label{table}
\begin{tabular}{lrrlllll}
\hline
 $U_1$ & $T_1$ &  $M_1$ & $\xH^-$ & $\delta(\xH^-)$ & $\varepsilon(\rho^-)$
       & $\varepsilon(\FLj)$ & $\varepsilon(\FR)$  \vbox to 9pt {}\\[1pt]
\hline
  65  &  8000  &   5.2  &  0.84  &  0.02  &  1.04   &  1.19   &  1.14  \\[2pt]  
  70  &  3000  &  10.9  &  0.47  &  0.01  &  1.02   &  1.08   &  1.03  \\
      &  6000  &   7.7  &  0.73  &  0.02  &  1.05   &  1.21   &  1.16  \\
      &  8000  &   5.6  &  1.    &  0.    &  1.08   &  1.30   &  1.28  \\[2pt]
  75  &  3000  &  11.7  &  0.62  &  0.02  &  1.03   &  1.13   &  1.07  \\  
      &  6000  &   8.2  &  0.75  &  0.04  &  1.04   &  1.20   &  1.14  \\  
      &  8000  &   6.0  &  1.    &  0.    &  1.10   &  1.33   &  1.30  \\[2pt]
  80  &  3000  &  12.5  &  0.86  &  0.05  &  1.09   &  1.38   &  1.32  \\
      &  6000  &   8.8  &  0.98  &  0.01  &  1.06   &  1.25   &  1.20  \\  
      &  8000  &   6.4  &  1.    &  0.    &  1.08   &  1.27   &  1.26  \\[2pt] 
  85  &  3000  &  13.2  &  1.    &  0.    &  1.10   &  1.30   &  1.26  \\
\hline
\end{tabular}
\end{table}

\begin{figure}
\resizebox{\hsize}{!}{\includegraphics{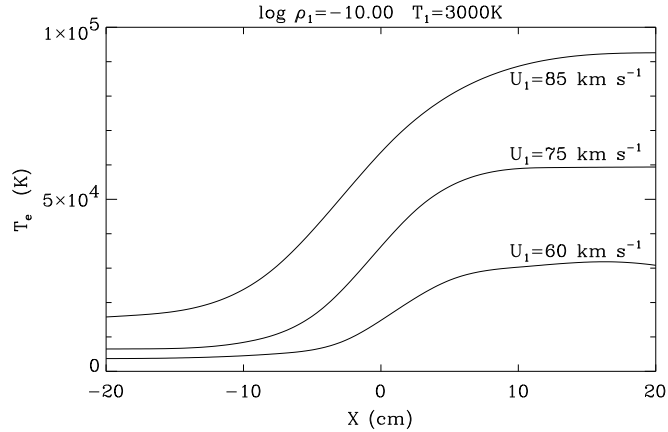}}
\caption{The electron temperature $\Te$ in the vicinity of the
discontinuous jump in radiative shock waves with electron thermal
conduction at upstream velocities $U_1 = 60$, 75 and $85~\kms$.}
\label{fig5}
\end{figure}

As is seen from Table~\ref{table},
the threshold upstream velocity is $U_1^*\approx 70~\kms$ and
effects of electron thermal conduction
become perceptible in strong shock waves with fully ionized hydrogen behind
the discontinuous jump.
At upstream gas flow velocities $U_1 \ge 70~\kms$ the postshock
electron temperature is enough high in order to support the necessary
temperature gradient and at the same to ionize more than a half
of hydrogen atoms in the radiative precursor
by the strong Lyman continuum flux emerging from the postshock region.
Shown in Fig.~\ref{fig5} plots of the electron temperature
in the vicinity of the discontinuous jump illustrate a very rapid
growth of the conductive precursor with increasing upstream
velocity $U_1$.
Indeed, in the upstream velocity range $70~\kms\le U_1\le 85~\kms$
the additional radiative losses due to effects of electron heat conduction
change from almost negligible values up to a quarter of the total radiative
flux.
Finally, as follows from Table~\ref{table},
the higher temperature of the ambient gas significantly favours
the electron heat conduction due to the higher
ionization of the unperturbed ambient gas.

\section{Conclusion}

In this paper we have shown that in radiative shock waves with
upstream velocities of
$U_1\gtrsim 70~\kms$ the physical properties of the gas surrounding
the discontinuous jump are sufficient for appearence of efficient
electron heat conduction, the conductive flux being comparable with the
Lyman continuum flux in the vicinity of the discontinuous jump.
The existence of the narrow conductive precursor affects all the region
of the Lyman continuum radiation trapped around the discontinuous jump
between zones of preshock ionization and postshock recombination
of the hydrogen gas.
As a consequence, the hydrogen ionization degree increases in the radiative
precursor by as much as 5\% provided that
the preshock hydrogen gas is partially ionized.
The effect of the saturation of the electron conductive flux
can be important only ahead the discontinuity but for models
considered in the present study we obtained $\vert\Fe\vert\lesssim\Fsat$.
The main conclusion of our study is that in shocks with velocities
exceeding the threshold value $U_1^*\approx 70~\kms$
the electron thermal conduction
significantly raises the radiative losses and, therefore, can
diminish the efficiency of the shock--driven mass loss.
More studies of radiative shock waves with electron heat
conduction for wider ranges of $\rho_1$, $T_1$ and $U_1$
are needed.

\begin{acknowledgements}
The work of YAF has been done in part under the auspices of the
Minist\`ere de la Recherche Fran\c caise as well as was supported
in part by the Russian National Program ``Astronomy'' (item 1102).
This paper is a part of PhD of H. Le Coroller.
\end{acknowledgements}

\end{document}